\documentclass[preprint,showpacs,preprintnumbers,amsmath,amssymb]{revtex4}

\usepackage{graphicx}
\usepackage{dcolumn}
\usepackage{bm}

\begin{document}

\newif\ifplot
\plottrue
\newcommand{\RR}[1]{[#1]}
\newcommand{\intsum}{\sum \kern -15pt \int}
\newfont{\Yfont}{cmti10 scaled 2074}
\newcommand{\Y}{\hbox{{\Yfont y}\phantom.}}
\def\O{{\cal O}}
\newcommand{\bra}[1]{\left< #1 \right| }
\newcommand{\braa}[1]{\left. \left< #1 \right| \right| }
\def\Bra#1#2{{\mbox{\vphantom{$\left< #2 \right|$}}}_{#1}
\kern -2.5pt \left< #2 \right| }
\def\Braa#1#2{{\mbox{\vphantom{$\left< #2 \right|$}}}_{#1}
\kern -2.5pt \left. \left< #2 \right| \right| }
\newcommand{\ket}[1]{\left| #1 \right> }
\newcommand{\kett}[1]{\left| \left| #1 \right> \right.}
\newcommand{\scal}[2]{\left< #1 \left| \mbox{\vphantom{$\left< #1 #2 \right|$}}
\right. #2 \right> }
\def\Scal#1#2#3{{\mbox{\vphantom{$\left<#2#3\right|$}}}_{#1}
{\left< #2 \left| \mbox{\vphantom{$\left<#2#3\right|$}}
\right. #3 \right> }}

\title{
Polarization Transfer Measurement for 
\mbox{\boldmath $^1{\rm H}(\vec{d},\vec{p})^2{\rm H}$}
Elastic Scattering at 135 MeV/u
and Three Nucleon Force Effects
}

\author{K.\ Sekiguchi$^{1}$}
\email{kimiko@rarfaxp.riken.go.jp}
\author{H.\ Sakai$^{1,2,3}$}
\author{H.\ Wita{\l}a$^{4}$}
\author{K.\ Ermisch$^{5}$}
\author{W.\ Gl\"ockle$^{6}$}
\author{J.\ Golak$^{4}$}
\author{M.\ Hatano$^{2}$}
\author{H.\ Kamada$^{7}$}
\author{N.\ Kalantar-Nayestanaki$^{5}$}
\author{H.\ Kato$^{2}$}
\author{Y.\ Maeda$^{2}$}
\author{J.\ Nishikawa$^{8}$}
\author{A.\ Nogga$^{9}$}
\author{T.\ Ohnishi$^{1}$}
\author{H.\ Okamura$^{10}$}
\author{T.\ Saito$^{2}$}
\author{N.\ Sakamoto$^{1}$}
\author{S.\ Sakoda$^{2}$}
\author{Y.\ Satou$^{11}$}
\author{K.\ Suda$^{8}$}
\author{A.\ Tamii$^{12}$}
\author{T.\ Uchigashima$^{2}$}
\author{T.\ Uesaka$^{3}$}
\author{T.\ Wakasa$^{13}$}
\author{K.\ Yako$^{2}$}
\affiliation{
$^{1}$RIKEN, the Institute of Physical and Chemical Research,
Wako, Saitama 351-0198, Japan}
\affiliation{
$^{2}$Department of Physics, University of Tokyo, Bunkyo,
Tokyo 113-0033, Japan}
\affiliation{
$^{3}$Center for Nuclear Study, University of Tokyo, Bunkyo,
Tokyo 113-0033, Japan}
\affiliation{
$^{4}$Institute of Physics, Jagiellonian University,
PL-30059 Cracow, Poland}
\affiliation{$^{5}$Kernfysisch Versneller Instituut, Zernikelaan
25, 9747 AA Groningen, The Netherlands}
\affiliation{
$^{6}$Institut f\"ur theoretische Physik II,
Ruhr-Universit\"at Bochum, D-44780 Bochum, Germany}
\affiliation{
$^{7}$Department of Physics, Faculty of Engineering,
Kyushu Institute of Technology, Kitakyushu 804-8550, Japan}
\affiliation{$^{8}$Department of Physics, Saitama University,
Saitama 338-8570, Japan}
\affiliation{$^{9}$Institute for Nuclear Theory, University of
Washington, Seattle WA 98195-1550, USA}
\affiliation{$^{10}$Cyclotron and Radioisotope Center, Tohoku
University, Sendai, Miyagi 980-8578, Japan}
\affiliation{$^{11}$Department of Physics, Tokyo Institute of
Technology
            Meguro, Tokyo 152-8551, Japan}
\affiliation{$^{12}$Research Center for Nuclear Physics, Osaka
University, Ibaraki, Osaka 567-0047, Japan}
\affiliation{$^{13}$Department of Physics, Kyushu University,
Fukuoka 812-8581, Japan}
%
%

\date{\today}

\begin{abstract}
The deuteron to proton polarization
transfer coefficients for the $d$--$p$ elastic scattering were
precisely measured with an incoming deuteron energy of 135 MeV/u 
at the RIKEN Accelerator Research Facility. The
data are compared to theoretical predictions based on exact
solutions of three-nucleon Faddeev equations with high--precision
nucleon--nucleon  forces combined with different 
three-nucleon forces (3NFs), 
representing the current, most popular models: the
$2\pi$-exchange Tucson-Melbourne model, a modification thereof
closer to chiral symmetry TM$'$(99), and the Urbana IX 3NF. 
Theory predicts large 3NF effects, especially in the 
angular range around the cross section minimum, but 
the present data only partially concurs, 
predominantly 
for $K_{xx}^{y'}-K_{yy}^{y'}$ ($K_{xx}^{y'}$, $K_{yy}^{y'}$)\@.
For the induced polarization, $P^{y'}$,
the TM$'$(99) and Urbana IX 3NFs reproduce
the data, but the Tucson-Melbourne 3NF
fails to describe the data.
For the polarization transfer coefficients, $K_{y}^{y'}$ and
$K_{xz}^{y'}$, the predicted 3NF effects 
are in
drastic conflict to the data. These facts clearly reveal the
defects of the 3NF models currently used.

\end{abstract}

\maketitle

\section{Introduction}

A main interest of nuclear physics is to
understand the forces 
acting between nuclear constituents.
Few nucleon systems offer 
decent opportunities to investigate these forces. 
Intensive theoretical and experimental efforts have
established high--precision nucleon--nucleon (NN)
potentials, partly based on one-meson exchange, partly on 
phenomenology, namely 
AV18~\cite{AV18}, CDBonn~\cite{cdb1,cdb2,cdb3}, 
Nijmegen I, II and 93~\cite{nijm}.
They reproduce a rich set of experimental 
NN data up to a laboratory energy 350 MeV with very high precision 
expressed in terms of 
a $\chi^2$ per data points very close to one. 
However, these so-called realistic NN
forces fail to predict the correct 
experimental binding energies of few-nucleon systems, but  
lead to clear underbinding.
For three- and four-nucleon systems, where exact solutions of 
the Schr\"odinger equation are available for these interactions, 
this underbinding 
amounts to 0.5 -- 1 MeV in case of $\rm ^3H$ and $\rm ^3He$, 
and to 2 -- 4 MeV for $\rm ^4He$~\cite{nogga}.
Also, for higher mass nuclei up to A=10, where stochastic
techniques have been applied, realistic NN forces 
fail to provide the measured binding energies~\cite{carl98,wir2000},
which is generally seen as the first hint to missing 
three-nucleon forces (3NFs) 
in the nuclear Hamiltonian.
Presently, the common 3NF 
models are based on a $2\pi$--exchange between three 
nucleons and the main ingredient in that process
is a $\Delta$--isobar excitation,
initially proposed by Fujita and Miyazawa 
almost half a century ago~\cite{fuj57}.
Further augmentations have led to 
the Tucson-Melbourne (TM)~\cite{tm3nf}
and the Urbana IX 3NF~\cite{urix}.\@ 
The TM 3NF was recently updated and now 
respects chiral symmetry, as noted in Refs.~\cite{fri99,huba1},  
and so we will also use the newest version from~\cite{tm99}, 
called TM$'$(99).
One can simultaneously achieve 
the correct binding energies for the  three-nucleon and 
four-nucleon systems by including the TM and Urbana IX 3NFs
into the nuclear Hamiltonian.
Also, adding the Urbana IX 3NF has successfully 
described the low energy bound states energies 
up to $\rm A=10$ nuclei.
Recently, it has been significantly improved by 
augmenting the Hamiltonian by the Illinois 3NF,
which are related to three-pion exchanges
with the intermediate $\Delta$'s~\cite{piep01}.

The binding energies of $s$-shell nuclei 
show the significance of 3NF,
but they only constrain its overall strength. 
The $p$-shell nuclei energies provide additional features.
In order to unambiguously clarify  
the detailed properties of 3NF at least for a total 
isospin of $\rm T = 1/2$,
the investigation of three nucleon scattering processes is required.
A rich set of energy dependent spin observables 
and differential cross sections are  
available in those reactions. 
Theoretical calculations based on 
several NN and 3N interaction models 
provide the theoretical guidance for selecting
specific observables and energies, which will 
appropriately determine the 3NF properties. 
The rapid progress in supercomputer technology 
has made it possible to achieve numerically exact 
solutions for the Faddeev equations up to an incident nucleon 
laboratory energy 
of 200 MeV using present day two-nucleon (2N) and three-nucleon (3N) 
potentials. 
The first clear signatures of the 3NF effects 
in the 3N continuum 
came from a study of minima in the differential cross section
for nucleon--deuteron ($Nd$) elastic scattering 
at incoming nucleon energies above $\approx~60$ MeV~\cite{wit98}. 
Including the $2\pi$-exchange TM 3NF in the nuclear Hamiltonian
removed a large part of the discrepancy between the data 
and theoretical predictions.
Calculations of the $Nd$ scattering 
in a coupled channel approach, when
$\Delta$-isobar degrees of freedom were explicitly included,
supported this conclusion~\cite{nem98}.
All these results confirm that 
the $Nd$ elastic scattering 
is a good tool for exploring the 3NF properties in this energy region.

The developments and progress 
in technology of highly polarized proton and deuteron  
ion sources and their application in recently 
constructed accelerators as well as 
new sophisticated techniques 
for target polarization, make it possible 
to obtain much more precise data for the 
spin observables 
at the higher energies ($E/A \gtrsim 60~\rm MeV$). 
Constructing highly efficient polarimeters
has also allowed accurate data on 
spin polarization transfer observables
to be obtained~\cite{sak2000,hatanaka02}. 
In Refs.~\cite{sak2000,sak96,sek2002} we have reported on
the precise data of the cross section and all deuteron
analyzing powers for $d$--$p$ elastic scattering
at incoming deuteron energies of 70, 100, and 135 MeV/u.
The data are compared with the theoretical
predictions based on various realistic NN potentials combined 
with different 3NFs, namely 
with the $2\pi$-exchange TM 3NF model, 
with a modification thereof closer to chiral symmetry 
(TM$'$)~\cite{fri99,huba1,kamada2001},
with the Urbana IX 3NF,
and with the phenomenological spin-orbit 3NF~\cite{kie99}.
For almost all observables, clear discrepancies
between the data and 2N force only predictions
are found especially in the cross section minima
and increase as the incident deuteron energy increases.
For the cross section, 
accounting for the 3NFs 
essentially removes these discrepancies.
For the deuteron vector analyzing power $A_y^d$, 
the 3NFs successfully explain 
the difference between the data and the 2N force only theoretical 
predictions.
Note that adding TM 3NF reproduces the 
recent data of $A_y^d$ and the spin correlation coefficient
$C_{y,y}$ at 197 MeV by Cadman {\it et al.}~\cite{cad01}.
However, 
theoretical predictions that incorporate 3NFs
(the TM, TM$'$ and Urbana IX 3NFs)
do not reproduce the deuteron tensor analyzing power data.
Recent proton vector analyzing power data have also revealed 
the deficiency of the 2$\pi$-exchange TM 3NF 
model~\cite{cad01,bie2000,steph99,erm01}\@
that yields large, incorrect effects.
The Urbana IX and TM$'$ 3NFs are 
much more successful and                                            %
provide a better description~\cite{erm01,wit01}.  

In the present study, 
we extend our measurement to new observables,
deuteron to proton ($\vec{d} + p \to \vec{p} + d $)
polarization transfer coefficients 
$K_{y}^{y'}$, $K_{xx}^{y'}-K_{yy}^{y'}$ ($K_{xx}^{y'}$, $K_{yy}^{y'}$),
and $K_{xz}^{y'}$ 
at 135 MeV/u 
in the region of the c.m. angles 
$\theta_{\rm c.m.} = 90^\circ - 180^\circ$.
These spin transfer coefficients are
predicted to have strong sensitivities to the 
current 3NF models~\cite{wit01}.
This is the first measurement of such polarization transfer
coefficients in this energy range ($E/A \gtrsim 60~\rm MeV$). 
To the best of our knowledge,
only proton to proton polarization transfer coefficients 
have been measured,
but at a much higher energy~\cite{hatanaka02}. 
The present data will provide 
a sensitive test for the 3NF models
in elastic $d$--$p$ scattering
below the pion production threshold energy.

In Section II the details of our experimental arrangement 
are presented. Section III provides a
description of the data analysis and experimental results.   
Section IV briefly reviews the basics of 
3N scattering formalism and gives a short description of the 3NFs   
used in this study. 
Our experimental results are compared with the theoretical predictions
in Section V\@, while Section VI\@ contains the summary and conclusion.

\section{Experimental Procedure}

\subsection{Polarized Deuteron Beams and Target}

The experiments were performed at the RIKEN Accelerator 
Research Facility (RARF) using the SMART system~\cite{ich94} 
including the focal plane polarimeter DPOL~\cite{ish95}.
The atomic beam type RIKEN polarized
ion source~\cite{oka94} provided 
the vector and tensor polarized deuteron beams. 
In the present measurements the data were taken with 
the vector and tensor polarization modes of the 
polarized and unpolarized deuteron beams given in terms
of the theoretical maximum polarization 
values as 
$(\it {\cal P}_Z,\it {\cal P}_{ZZ}) = \rm (0,0)$, $(0,-2)$, 
$(-2/3,0)$ and $(1/3,1)$.
These polarization modes were cycled in 5--second intervals
by switching the RF transition units of the ion source. 
The deuteron polarization axis was rotated by a spin rotation system 
Wien Filter~\cite{oka95} prior to acceleration.
It was perpendicular to the scattering plane when
measuring $K_y^{y'}$ and $K_{yy}^{y'}$. 
For $K_{xx}^{y'}$, the rotation was performed into the
scattering plane so that the polarization axis pointed sideways,
perpendicular to the beam.
For the $K_{xz}^{y'}$ measurement, 
the spin symmetry axis was additionally 
rotated in the reaction plane and 
aligned at an angle $\beta$ to the beam direction.
A typical value of $\beta$ was $131.6^\circ \pm 0.2^\circ$.
The beam polarization was monitored by 
$d$--$p$ elastic scattering at 135 MeV/u 
and it was $60$ -- $80$\% of 
the theoretical maximum values throughout the measurement. 
Polyethylene ($\rm CH_2$) with a thickness of $90$-$\rm mg/cm^2$
or the liquid hydrogen with a thickness of $20$-$\rm mg/cm^2$,
employed as a hydrogen target ($\rm ^1H$)~\cite{ues2001},
was bombarded with a beam intensity of $10$ -- $60$ nA.

\subsection{Beam Line Polarimeter}

 Two sets of beam line polarimeters monitored  
 the beam polarization. 
 The first, the D--room polarimeter, was installed downstream of
 the Ring cyclotron, which was used to determine the beam
 polarization after the deuterons were accelerated by the Ring
 cyclotron.
 The second, the Swinger polarimeter (see Fig.\ {1}), 
 was placed in front         
 of the scattering chamber in the experimental room.
 Since the incident beam direction was rotated 
 using the beam-swinger system of the SMART,
 the polarization axis of the beam was precessed 
 during the beam transportation from the D--room polarimeter
 to the target position.
 The Swinger polarimeter moved with the beam swinger 
 so that this polarimeter could directly measure the beam
 polarization at the target.
 The beam polarization before and after 
 each run was measured using the Swinger polarimeter.

 The polarimetry was made by using the analyzing powers for 
 $d$--$p$ elastic scattering.
 To obtain the absolute values of the deuteron beam polarizations, 
 the analyzing powers for $d$--$p$ elastic 
 scattering were calibrated by using the 
 $^{12}{\rm C}(d,\alpha)^{10}{\rm B}^{*}\left[2^{+}\right]$
 reaction,
 the $A_{yy}\left(0^\circ\right)$ of which 
 is exactly $-1/2$ because of parity conservation~\cite{suda2002}.
A $\rm CH_2$ sheet was the target for each polarimeter.
The target thickness was
$270$-$\rm mg/cm^2$ for the D-room polarimeter
and $90$-$\rm mg/cm^2$ for the Swinger polarimeter. 
Each polarimeter consisted of four pairs of $\rm 1$-cm thick 
plastic scintillators placed symmetrically 
in left, right, up and down directions. 
The scattered deuterons and recoil protons were
detected in a kinematical coincidence. This setup reduced 
background events due to the deuteron breakup 
process or the inelastic scattering from carbon nuclei.

\subsection{SMART system and focal plane polarimeter DPOL}

The polarization transfer measurement was performed using
the SMART system~\cite{ich94} with the focal plane polarimeter 
DPOL~\cite{ish95} (see Fig.~\ref{smart}).
The polarized deuteron beam bombarded the hydrogen target
placed in the scattering chamber. Recoil protons were
momentum analyzed by the magnetic spectrograph 
and detected at its second focal plane (FP2 in Fig.~\ref{smart}). 
In the SMART system, the magnetic spectrograph was fixed
to the ground and the incident beam direction was rotated
by the Swinger magnet, leading to a vertical reaction plane.

The FP2 detector system consisted of 
a multi-wire drift chamber (MWDC1 in Fig.~\ref{dpol})
and three plastic scintillation counters (SC1--3 in Fig.~\ref{dpol}).
The MWDC1 was used to reconstruct 
the trajectories of the particles at the FP2.
The configuration of the wire planes 
was X--Y--X$'$--Y$'$--X$'$--Y$'$--X--Y 
and the coordinate frames were defined as follows.
The ${\textit {\textbf z}}$--axis referred to the central ray.
The ${\textit {\textbf x}}$--axis was perpendicular to the 
${\textit {\textbf z}}$--axis
in the horizontal plane and 
the ${\textit {\textbf y}}$--axis was taken as 
$\hat{{\textit{\textbf x}}} \times \hat{{\textit{\textbf z}}}$.
All position sensitive planes were normal 
to the ${\textit {\textbf z}}$--axis
and separated by a distance of 50 mm from 
adjacent planes. 
The planes with primes were displaced half a cell relative to
the unprimed ones which helped solve the so called 
left-right ambiguity.
The cell size was 20 mm $\times$ 20 mm for the X--planes and 
10 mm $\times$ 10 mm for the Y--planes.
The plastic scintillation counters BICRON BC-408 
of the size 180 mm$^{H} \times$ 800 mm$^{W} \times$ 5 mm$^{T}$
(SC1--3 in Fig.~\ref{dpol}) were used to identify 
proton events scattered on the hydrogen target
and to generate event triggers.
The photo-multiplier tubes Hamamatsu H1161 
were placed at both ends of 
the scintillators via light guides.

Proton polarization was measured by the DPOL
after momentum analysis in the magnetic spectrograph.
The DPOL consisted of an analyzer target, 
a multiwire drift chamber (MWDC2 in Fig.~\ref{dpol}),
and the counter hodoscope system (HOD and CM in Fig.~\ref{dpol}).
The DPOL was primarily designed and 
optimized for the deuteron polarization measurements
and then was modified to measure the proton polarization. 

The polarimetry was made using the $p + \rm C$ scattering. 
As an analyzer target, a 3-cm thick carbon plate was 
sandwiched between the two plastic 
scintillation trigger counters (SC1 and SC2 in Fig.~\ref{dpol}).
The trajectories of the scattered protons from 
the $p + \rm C$ reaction
were reconstructed by the MWDC2\@.
The MWDC2 was $\rm 670~mm$ downstream 
from the exit window of the MWDC1 and had eight layers of 
sense wire planes with the 
$\rm Y_f$-$\rm Y_f'$-$\rm X_f$-$\rm X_f'$-$
\rm Y_r$-$\rm Y'_r$-$\rm X_r$-$\rm X'_r$ configuration.
Here ``f'' and ``r'' denote the front and rear planes, respectively.
The planes with primes were, again, displaced half a cell 
relative to the unprimed ones.
The coordinate frame was defined as in the case of the MWDC1\@.
The cell size was 
$\rm 14~mm~\times 14~mm$ ($\rm 15~mm~\times 15~mm$)
for the $\rm X_f$ ($\rm X_r$ ) planes
and 
$\rm 14~mm~\times 14~mm$ ($\rm 16~mm~\times 16~mm$)
for the $\rm Y_f$ ($\rm Y_r$) planes.
The number of cells was 64 for the X--planes and 32 for the Y--planes.

Event triggers for proton events from the $p + \rm C$ reaction
were generated by making a coincidence
of the signals of the SC1--3 counters and
those of the counter hodoscope system which was
located $\rm 4~m$ downstream from the analyzer target. 
The angular range covered by the hodoscope system
was $\pm 15^{\circ}$ both vertically and horizontally.
The unscattered protons
passed through insensitive region 
between the upper and lower parts of the hodoscope.
The front wall of the hodoscope (HOD in Fig.~\ref{dpol}) was comprised of
a layer of 28 segmented plastic scintillators, which were
$\rm 2200$-mm wide, $\rm 65$-mm high and $\rm 65$-mm thick.
The light output signals from each HOD 
were read out by two photo-multiplier tubes Hamamatsu H1161 
that were directly coupled to each scintillator at both ends.
The rear wall consisted of six plastic scintillators 
(CM in Fig.~\ref{dpol}).
Each CM counter was $\rm 2200$-mm wide, $\rm 190$-mm high and 
$\rm 10$-mm thick. 
Two photo multiplier tubes Hamamatsu H1161
were attached to both ends of 
each scintillation counter via light guides. 
In the angular range $\theta_{\rm c.m.} = 120^\circ$--$180^\circ$
for $d$--$p$ elastic scattering,
the CM counters were used to 
generate the $p + \rm C$ event triggers
by taking the coincidence with the HOD and SC1--3 signals.
However, for angles $\theta_{\rm c.m.} = 90^\circ$--$110^\circ$
the kinetic energies of the scattered protons were too low 
to allow them to reach the CM counters. 
Under these circumstances, 
the CM signals were not used as the event trigger.

Data acquisition was carried out
with a fast data acquisition 
system for the SMART spectrograph~\cite{oka2000}.
The data were accumulated 
in a VME memory module through the FERA bus
and then DMA-transferred to a personal computer.

\section{Data Analysis and Experimental Results}

\subsection{Polarization Transfer Coefficients}

\subsubsection{Coordinate Frame for the Polarization Observables at the SMART System}

The coordinate frame for the polarization observables 
in the SMART system is defined according 
to the Madison convention \cite{ohl72},
shown in Fig.~\ref{smartconfig}.
The $z$-axis is given by the beam direction.
The $y$-axis is perpendicular to the reaction plane and 
the $x$-axis is defined by
$\hat{{\textit{\textbf y}}} \times \hat{\textit {\textbf z}}$.
The coordinate system $\left( x',y',z' \right)$ for the 
polarization of the scattered protons is rotated through the 
dipole magnet of the SMART spectrograph 
into the coordinate system at FP2 $(x'',y'',z'')$.

In Fig.~\ref{smartconfig}, the $\{p_{ij}\}$ 
is the vector or tensor deuteron beam polarizations,
the $p_{i'}$ is the polarization of
the scattered protons and 
the $p_{i''}$ is the polarization of 
the scattered protons at FP2\@.
In the present measurement, the $p_{y''}$ was measured with 
the focal plane polarimeter DPOL and the $p_{y'}$ 
was extracted using the $p_{y''}$ 
and the spin precession angle $\chi$
in the dipole magnets of the spectrometer (see Sec.~III.A.2).

\subsubsection{Effective Analyzing Power Measurement}

The effective analyzing powers $A_y^C$ 
of the DPOL were calibrated
at three proton energies 120, 144 and 200 MeV
which almost covered the kinetic energy region
of scattered protons for $d$--$p$ elastic scattering 
($E_p^{\rm scatt.}=120-240\rm MeV$).
Since the polarized proton beams were unavailable at RARF,
the induced polarization $P^{y'}$ in the 
$^{12}{\rm C}(p,\vec{p})^{12}{\rm C}$ elastic scattering
was used to determine the $A_y^{C}$.
The analyzing powers $A_y$ for the time-reversed reaction  
$^{12}{\rm C}(\vec{p},p)^{12}{\rm C}$ 
are equal to the induced polarizations $P^{y'}$ and
were precisely measured at $E_p = $ 122 and 200 MeV
by Meyer {\it et al.}\ at IUCF~\cite{1981Me02,1983Me02}.
For 200 MeV, 
the calibrations were performed by using the two 
spin modes of the polarized proton beams 
obtained 
by the $^{12}{\rm C}(p,\vec{p})^{12}{\rm C}$ elastic scattering
at $\theta_{\rm lab.} = 16.1^\circ$ and $28.1^\circ$.
The expected values of the polarizations $P^{y'}$ were 
$0.993$ and $-0.425$ for the $\theta_{\rm lab.} = 16.1^\circ$ 
and $28.1^\circ$, respectively.
For 120 MeV,
the calibration was performed at the angle 
$\theta_{\rm lab.} = 24.2^\circ$ where 
the polarization of the proton beams was
expected to be $0.715$.
Passing 200 MeV proton beams through a brass plate, 
which was in front of the MWDC1 
just downstream of the exit window of the D2 magnet,
reduced the energy to create 144 MeV polarized proton beams.
A $284$-$\rm mg/cm^2$ thick graphite target 
in the SMART scattering chamber was bombarded 
by unpolarized proton beams and 
the scattered protons bombarded the polarization analyzer target.
Since the $y''$-axis is in the horizontal plane
in the SMART system (see Fig.~\ref{smartconfig}), 
the up-down asymmetry 
was used to extract the effective analyzing power $A_y^{C}$. 
The $A_y^{C}$ is given by
\begin{eqnarray}
A_y^{C} &=&
 \frac{\int_{-\Delta \phi}^{\Delta \phi}
 \int_{\theta_{\rm min}}^{\theta_{\rm max}}
 I_{\rm o}\left(\theta\right) A_y\left(\theta \right)
 \sin \theta \cos \phi
 d\theta d\phi}
 {\int_{-\Delta \phi}^{\Delta \phi}
 \int_{\theta_{\rm min}}^{\theta_{\rm max}}
 I_{\rm o} \left(\theta \right)
 \sin \theta  d\theta d\phi}
 \label{aycdef}. 
\end{eqnarray}
The numbers of events in the upper, $N_U$, and
lower, $N_D$, side region are obtained as,
\begin{eqnarray}
N_U &=& 
 \int_{-\Delta \phi}^{\Delta \phi}
 \int_{\theta_{\rm min}}^{\theta_{\rm max}}
 I_{\rm o}\left(\theta\right)
 \left[ 1 + A_y\left(\theta \right) p_{y''} \cos \phi \right] \sin \theta 
 d\theta d\phi \label{NU}, \\
N_D &=& 
 \int_{-\Delta \phi + \pi }^{\Delta \phi + \pi}
 \int_{\theta_{\rm min}}^{\theta_{\rm max}}
 I_{\rm o}\left(\theta\right) 
 \left[ 1 + A_y\left(\theta \right) p_{y''} \cos \phi \right] \sin \theta 
 d\theta d\phi \label{ND}.
\end{eqnarray}
Here, the $I_{\rm o}\left(\theta\right)$ and $A_y\left(\theta\right)$
are the cross section and the analyzing power for inclusive proton
scattering in the carbon analyzer of the DPOL.
The $p_{y''}$ is the proton beam polarizations at the FP2 shown 
in Fig.~\ref{smartconfig}.
To reduce the instrumental asymmetries, 
the $A_y^{C}$ was extracted in the 
following way.
From Eqs.~(\ref{aycdef}), (\ref{NU}), and (\ref{ND})
the $N_U$ and $N_D$ were normalized as,
\begin{eqnarray}
n_U &\equiv& \frac{N_U}{N_{U}+N_{D}} = 
 \frac{1}{2}\ \left( 1 + A_y^C p_{y''} \right), \\
n_D &\equiv& \frac{N_D}{N_{U}+N_{D}} = 
 \frac{1}{2}\ \left( 1 - A_y^C p_{y''} \right).
\end{eqnarray}
The spin-up ($p_{y''} ^{+}$; $p_{y''} > 0$) and spin-down 
($p_{y''} ^{-}$; $p_{y''} < 0$) 
polarized proton beams used in the measurement 
together with the corresponding $n_U^\pm$ and $n_D^\pm$ provide
the effective analyzing power $A_y^C$ as,
\begin{eqnarray}
A_y^C &=& \frac{\left( n_U^{+} - n_D^{+} \right) 
 - \left( n_U^{-} - n_D^{-}\right)}  
             {p_{y''}^{+} - p_{y''}^{-}}\ .
\end{eqnarray}
In the 120MeV measurements, 
the data were collected with the one-mode polarized 
proton beams. 
Therefore in analysis we also used 
the data with unpolarized beams, 
which was obtained by directly tuning incident proton beams 
to the polarization analyzer target at the focal plane.
Angular integrations in Eqs.~(\ref{NU}) and (\ref{ND})
were performed 
over regions of polar and azimuthal 
angles of $5^\circ \le \theta \le 15^\circ$ and
$\Delta \phi = 60 ^\circ$, respectively.
The proton spin precessed around the vertical axis of the 
spectrometer and the spin precession angle $\chi$ with
respect to the direction of the proton momentum is 
given in the moving frame 
by $\chi = \gamma\left(g/2 - 1 \right)\Theta_{D}$,
where $\gamma$ is the Lorentz 
factor $\gamma=\left(m_pc^2 + E_p \right)/m_pc^2$,
$g$ is the spin $g$ factor of the proton, 
and $\Theta_{D}$ is the bending angle of the spectrometer.
The total bending angle of the magnetic spectrograph is
$\Theta_{D} = 60^\circ$. Thus 
the $p_{y''}$ is given as,
\begin{eqnarray}
p_{y''}&=& P^{y'}\cos \chi.
\label{lorentz}
\end{eqnarray}
Figure~\ref{aycfit} shows the measured effective 
analyzing power $A_y^C$ with open circles as 
a function of the proton energy at the center of the 
carbon plate $E_p^{C}$.
Only the statistical errors are shown.
The $E_p^{C}$ was calculated by numerically
integrating the energy loss per unit thickness described 
by the Bethe-Bloch equations.
The energy dependent curve of the $A_y^C$
was obtained by fitting the effective analyzing powers
calculated from the empirical-energy-dependent fit of
the inclusive analyzing powers for the $p + \rm C$ 
by McNaughton {\it et al.}~\cite{MCN85} and 
the angular distributions of the differential cross section
of Aprile-Giboni {\it et al.}~\cite{apl83}.
The obtained curve was scaled to adjust the experimentally
obtained $A_y^C$ (a dotted curve in Fig.~\ref{aycfit}).
The uncertainty of the input parameters for the $p + \rm C$ 
inclusive analyzing power~\cite{MCN85} is $2\%$.
The uncertainty of the fit for the energy dependent curve 
is 6\%, in which the scaling factor has an uncertainty of $2\%$.
Thus, the estimated overall systematic uncertainty of the 
effective analyzing power, $A_y^C$, of the DPOL is $7\%$.

\subsubsection{Extraction of the Polarization Observables
\label{ext_pt}}

The polarization transfer coefficients 
for the reaction 
$\vec{d} + p \to \vec{p} + d $
are expressed through 
the unpolarized $(\sigma_0)$ and polarized $(\sigma)$ 
cross sections together with 
the polarizations of incoming deuteron $(p_{ij})$ and
outgoing protons $(p_{k'})$ as, 
\begin{eqnarray}
p_{x'} \sigma/\sigma_0 &=& 
 \frac{3}{2}p_x K_x^{x'} + 
 \frac{3}{2}p_z K_z^{x'} + 
 \frac{2}{3}p_{xy} K_{xy}^{x'} +  
 \frac{2}{3}p_{yz} K_{yz}^{x'}, \label{kij_x}\\
p_{y'} \sigma/\sigma_0 &=& 
 P^{y'} +
 \frac{3}{2}p_y K_y^{y'} + \frac{2}{3}p_{xz} K_{xz}^{y'}  +
 \frac{1}{3}
 \left(p_{xx} K_{xx}^{y'} +  
       p_{yy} K_{yy}^{y'} +  
       p_{zz} K_{zz}^{y'}\right),  \label{kij_y}\\
p_{z'} \sigma/\sigma_0 &=& 
 \frac{3}{2}p_x K_x^{z'} + 
 \frac{3}{2}p_z K_z^{z'} + 
 \frac{2}{3}p_{xy} K_{xy}^{z'} +  
 \frac{2}{3}p_{yz} K_{yz}^{z'}, \label{kij_z}
\end{eqnarray}
where $x$, $y$, and $z$ ($x'$, $y'$, and $z'$)
are the coordinate system used to describe polarizations 
of the incident deuterons (outgoing protons) \cite{ohl72}.

To extract the polarization transfer coefficients
$K_{y}^{y'}$, $K_{xx}^{y'}$, $K_{yy}^{y'}$, and $K_{xz}^{y'}$,
we applied the polarized deuteron beams with
the spin symmetry axis 
directed in the optimum orientation 
for each observable.
We rotated it to the $y$-axis
for the $K_y^{y'}$ and $K_{yy}^{y^\prime}$, and
to the $x$-axis for the $K_{xx}^{y^\prime}$ measurement.
For the $K_{xz}^{y^\prime}$ measurement 
we rotated the spin symmetry axis into the reaction plane
and additionally aligned it at the angle $\beta$ to
the beam direction ($\beta = 131.6^\circ \pm 0.2^\circ$).
To obtain $K_{xx}^{y'} - K_{yy}^{y'}$,
$K_{xx}^{y'}$ and $K_{yy}^{y'}$ were independently measured.
Accordingly the polarized cross section can be written 
for each observable as, 
\begin{eqnarray}
p_{y'} \sigma/\sigma_0 &=& 
 P^{y'} +
 \frac{3}{2}p_y K_y^{y'} + \frac{1}{2}p_{yy} K_{yy}^{y'} 
 ~~~~~~~~~~~~\ {\rm for}\ K_{y}^{y'}\ {\rm and } \ K_{yy}^{y'}, 
 \label{shiki_kij1}\\
p_{y'} \sigma/\sigma_0 &=& 
 P^{y'} +
 \frac{1}{2}p_{xx} K_{xx}^{y'}  
 ~~~~~~~~~~~~~~~~~~~~~~~~~~~~~~~~~~~~\ {\rm for } \ K_{xx}^{y'}, 
 \label{shiki_kij2}\\
p_{y'} \sigma/\sigma_0 &=& 
 P^{y'} +
 \frac{2}{3}p_{xz} K_{xz}^{y'} \nonumber \\
&&+  \frac{1}{3}\left(p_{xx} - p_{zz} \right) K_{xx}^{y'}   
  + \frac{1}{3}\left(p_{yy} - p_{zz} \right) K_{yy}^{y'} 
  ~~~~~~\ {\rm for } \ K_{xz}^{y'} 
  \label{shiki_kij3}\\
 && {\rm with~}~K_{xx}^{y'} + K_{yy}^{y'} + K_{zz}^{y'} = 0. \nonumber
\end{eqnarray}

By using the relation between 
the deuteron beam polarizations $({\cal P}_Z,{\cal P}_{ZZ})$ 
(see Sec.~II.A) and $(p_i,p_{lk})$ 
given by the angles $(\beta,\phi)$~\cite{ohl72} and 
the relation 
$p_{y''} = p_{y'}\cos \chi$ (see Sec.~III.A.2),
the polarized cross section for each polarization transfer 
coefficient given in Eqs.~(\ref{shiki_kij1})--(\ref{shiki_kij3})
is expressed as, 
\begin{eqnarray}
p_{y''} \sigma/\sigma_0 &=& 
p_{y'}  \sigma/\sigma_0 \cos \chi
\nonumber \\
&=&(P^{y'} + {v}_e {\cal P}_{Z} + {t}_e {\cal P}_{ZZ}) \cos \chi 
\nonumber \\
&=& {\cal P}^{y'} + {\cal V}_e {\cal P}_{Z} + {\cal T}_{e} {\cal P}_{ZZ},
\end{eqnarray}
where 
\begin{eqnarray}
{\cal P}^{y'} &=& P^{y'}\cos\chi.
\end{eqnarray}
For $K_y^{y'}$ and $K_{yy}^{y'}$, the ${\cal V}_e$ and ${\cal T}_e$ are
given as, 
\begin{eqnarray}
{\cal V}_e &=& \frac{3}{2}K_y^{y'} \cos \phi \sin \beta \cos \chi, \\
{\cal T}_e &=& \frac{1}{2} K_{yy}^{y'} 
   \left( \sin^2\beta\cos^2\phi - \cos^2\beta \right) \cos\chi \nonumber \\
   &=& \frac{1}{2}K_{yy}^{y'} \cos \chi, 
\end{eqnarray}
where $\left(\beta,\phi \right) = \left(90^\circ,0^\circ\right)$.\\
\indent
For $K_{xx}^{y'}$,
\begin{eqnarray}
{\cal T}_e &=& 
        \frac{1}{2} K_{xx}^{y'} \left( \sin^2\beta\sin^2\phi - \cos^2\beta \right)
        \cos\chi \nonumber \\
 &=& \frac{1}{2} K_{xx}^{y'}\cos \chi,
\end{eqnarray}
where $\left(\beta,\phi \right) = \left(90^\circ,-90^\circ\right)$.\\
\indent
For  $K_{xz}^{y'}$,
\begin{eqnarray} 
{\cal T}_e &=& 
	\Bigl\{- K_{xz}^{y'} \sin\beta\cos\beta\sin\phi 
	+ \frac{1}{2}K_{xx}^{y'}
	\left( \sin^2\beta\sin^2\phi - \cos^2\beta \right) \nonumber \\
 && + \frac{1}{2}K_{yy}^{y'}
    \left( \sin^2\beta\cos^2\phi - \cos^2\beta \right) \Bigr\} \cos \chi.
\end{eqnarray}
It should be noted that the $K_{xz}^{y'}$ value was extracted
using the measured $K_{xx}^{y'}$ and $K_{yy}^{y'}$ values. 

The $p_{y''}$ is obtained from the Eqs.~(\ref{NU}) and (\ref{ND}) as,
\begin{eqnarray}
p_{y''} 
&=& \frac{A_{\rm sym.}}{A_y^C}, \\
A_{\rm sym.} & \equiv & \frac{N_U - N_D}{N_U + N_D}  \nonumber.
\label{asymmetry}
\end{eqnarray}
From the resulting values of $p_{y''}^{(i)}$ for each spin-mode
\#i (see Sec.~II.A), 
the ${\cal P}^{y'}$, ${\cal V}_e$  and ${\cal T}_e$ 
values were calculated as,
\begin{eqnarray}
{\cal P}^{y'}_{[1]} &=& p_{y''}^{(0)}\ ,\label{induc_1} \\
{\cal P}^{y'}_{[2]} &=& 
\biggl\{
 p_{y''}^{(1)}R^{(1)}\left({\cal P}_{Z}^{(2)}{\cal P}_{ZZ}^{(3)}
		          -{\cal P}_{Z}^{(3)}{\cal P}_{ZZ}^{(2)}\right)
+p_{y''}^{(2)}R^{(2)}\left({\cal P}_{Z}^{(3)}{\cal P}_{ZZ}^{(1)}
		          -{\cal P}_{Z}^{(1)}{\cal P}_{ZZ}^{(3)}\right)
\nonumber \\
&&
+p_{y''}^{(3)}R^{(3)}\left({\cal P}_{Z}^{(1)}{\cal P}_{ZZ}^{(2)}
		          -{\cal P}_{Z}^{(1)}{\cal P}_{ZZ}^{(2)}\right)
\biggr\} \nonumber \\
&&
\bigg/
\biggl\{
 \left({\cal P}_{Z}^{(2)}{\cal P}_{ZZ}^{(3)}-{\cal P}_{Z}^{(3)}{\cal P}_{ZZ}^{(2)}\right)
+\left({\cal P}_{Z}^{(3)}{\cal P}_{ZZ}^{(1)}-{\cal P}_{Z}^{(1)}{\cal P}_{ZZ}^{(3)}\right)
+\left({\cal P}_{Z}^{(1)}{\cal P}_{ZZ}^{(2)}-{\cal P}_{Z}^{(1)}{\cal P}_{ZZ}^{(2)}\right)
\biggr\}
, \nonumber \\
\label{induc_2}\\
{\cal V}_e  &=& 
\Biggl[
p_{y''}^{(2)} - p_{y''}^{(3)}  - {\cal A}_T\ p_{y''}^{(1)}-
\biggl\{ 
\frac{1}{R^{(2)}} - \frac{1}{R^{(3)}} - {\cal A}_T\ \frac{1}{R^{(1)}}
\biggr\}\
p_{y''}^{(0)}\ \Biggr] \nonumber \\
&&
\bigg/
\biggl\{
 \frac{{\cal P}_Z^{(2)}} {R^{(2)}} - \frac{{\cal P}_Z^{(3)}} {R^{(3)}}
-{\cal A}_T \
 \frac{{\cal P}_Z^{(1)}} {R^{(1)}}
\biggr\}, \label{ky-y}
\\
\nonumber \\
{\cal T}_e  &=& 
\Biggl[
p_{y''}^{(1)} - p_{y''}^{(3)}  - {\cal A}_V\ p_{y''}^{(2)}-
\biggl\{ 
\frac{1}{R^{(1)}} - \frac{1}{R^{(3)}} -  {\cal A}_V\ \frac{1}{R^{(2)}}
\biggr\}\
p_{y''}^{(0)}\ \Biggr] \nonumber \\
&&
\bigg/
\biggl\{
 \frac{{\cal P}_{ZZ}^{(1)}} {R^{(1)}} - \frac{{\cal P}_{ZZ}^{(3)}} {R^{(3)}}
-{\cal A}_V \
 \frac{{\cal P}_{ZZ}^{(2)}} {R^{(2)}}
\biggr\},\label{kyy-y}
\end{eqnarray}
where
\begin{eqnarray}
{\cal A}_V &=& \frac{R^{(2)}}{{\cal P}_Z^{(2)}}\
 \biggl\{ \frac{{\cal P}_Z^{(1)}} {R^{(1)}}
        -\frac{{\cal P}_Z^{(3)}} {R^{(3)}}
 \biggr\}, \\
{\cal A}_T &=& \frac{R^{(1)}}{{\cal P}_{ZZ}^{(1)}}\
 \biggl\{ \frac{{\cal P}_{ZZ}^{(2)}} {R^{(2)}}
        -\frac{{\cal P}_{ZZ}^{(3)}} {R^{(3)}}
 \biggr\},
\end{eqnarray}
with $R^{(i)}  =  \sigma ^{(i)} / \sigma_{\rm o} $.

The induced polarization was obtained 
using ${\cal P}^{y'}_{[1]}$, ${\cal P}^{y'}_{[2]}$ 
in Eqs.~(\ref{induc_1}) and (\ref{induc_2}), 
respectively, and the resulting values 
were consistent with each other within statistical accuracy.
Finally, the $P^{y'}_{[1]}$ and $P^{y'}_{[2]}$ averaged
with the statistical weights were used to minimize the errors
when determining the $P^{y'}$ value.

Figure~\ref{ex_LiQ} shows the excitation energy spectra
at the angles $\theta_{\rm c.m.} = 176.8^\circ$,
$120.0^\circ$, and $90.0^\circ$ 
obtained with the liquid hydrogen target.
At $\theta_{\rm c.m.} = 176.8^\circ$, 
the portion of the spectrum due to  
final state interaction (FSI) of $d$--$p$ breakup reaction 
is clearly seen at the energies $E_{x} \gtrsim 2~\rm MeV$
and it is well separated from $d$--$p$ elastic 
scattering events. The kinetic energy of the outgoing proton 
for $d$--$p$ elastic scattering 
changes rapidly with the scattering angle and 
loses the energy resolution 
at forward angles in c.m.\@. 
Therefore spectra due to elastic scattering
and breakup reactions are not clearly separated at the angles 
$\theta_{\rm c.m.} \le 140^\circ$
(see the spectra for the angles 
$\theta_{\rm c.m.} = 120^\circ$ and $90^\circ$ in Fig.~\ref{ex_LiQ}).
To reduce the background, only events in the hatched region
were selected to obtain the 
polarization observables for $d$--$p$ elastic scattering.
The position of the hatched energy region did not 
include the energy region $E_{x}\ge 2~\rm MeV$,
the region that clearly included the breakup reactions. 
To see the background contributions in the energies 
$E_{x} \le 2~\rm MeV$, the polarization observables were 
obtained by changing the maximum energy value in the 
hatched energy region. 
The magnitude of polarization observables changed 
by 0.02 or less.
Typically, 
an integration range $E_{x} \le 0.5~\rm MeV$ 
was adopted to extract final polarization observables.

The experimental results for the polarization transfer coefficients
($K_{y}^{y'}$, $K_{yy}^{y'}$, $K_{xx}^{y'}$, and $K_{xz}^{y'}$)
and the induced proton polarization $P^{y'}$ 
are shown with open circles in Fig.~\ref{kij}
and tabulated in Table~\ref{kij_tab}.
Only the statistical uncertainties  are shown and
their magnitudes are less than
$0.02$ for $P^{y'}$
and less than
$0.03$ for all polarization transfer coefficients
($K_{y}^{y'}$, $K_{yy}^{y'}$, $K_{xx}^{y'}$, $K_{xz}^{y'}$).

The fluctuation of the polarization transfer coefficients
from the uncertainty of the bending angle of the spectrometer
is less than $1\%$.
The uncertainty of the 
effective analyzing power for the DPOL is $7 \%$.
The deuteron beam polarizations have an 
uncertainty of less than $3 \%$.
The effects of the breakup reactions at the angles 
$\theta_{\rm c.m.} \le 140~\rm MeV$,
in which the events were inseparable from the elastic ones,
were 0.02 or less in magnitude and then the systematic
uncertainties coming from the breakup reactions 
did not override the statistical ones.
Thus the overall systematic uncertainties 
are estimated to be about $8\%$
for the polarization transfer coefficients and the 
induced polarization $P^{y'}$.
For the induced polarization $P^{y'}$, 
our data were compared with the proton 
analyzing power $A_y^{p}$ for $p$--$d$ elastic scattering
measured at KVI~\cite{erm01} (solid squares in Fig.~\ref{kij}).
Assuming time-reversal invariance $P^{y'} = -\ A_y^{p}$,
and these two independent measurements agree with each other
within the statistical uncertainties in the measured angular
range, $\theta_{\rm c.m.} = 90^\circ - 180^\circ$.
Figure~\ref{kij} shows
the data obtained in the test measurement~\cite{sak2000}
(open circles) together with the present data.
These two measurements are consistent, except for
$K_{yy}^{y'}$ at $\theta_{\rm c.m.} = 150^\circ$.

\subsection{Analyzing Powers}

As described in Sec.~II.B,
the analyzing powers for $d$--$p$ elastic 
scattering were used to obtain the deuteron 
beam polarizations.
Recently, 
to determine the absolute values 
of beam polarizations, the analyzing powers 
for $d$--$p$ elastic scattering 
were calibrated at six angles 
for deuteron energies of  70 and 135 MeV/u,
by using the reaction 
$^{12}{\rm C}(d,\alpha)^{10}{\rm B}^{*}\left[2^{+}\right]$
at $0^\circ$ \cite{suda2002}\@. 
Tables \ref{aij_suda70} and \ref{aij_suda135} show the data.
The previously reported data in Ref.~\cite{sek2002}
were not extracted with these new calibration data 
but with those obtained 
using the $^{12}{\rm C}(\vec{d},p)^{13}{\rm C}$
reaction or $^{3}{\rm He}(\vec{d},p)^{4}{\rm He}$
reaction at low energies~\cite{sak96,ues01}.
In the analysis in Ref.~\cite{sek2002}, 
the analyzing power data at $\theta_{\rm c.m.} = 90.0^\circ$ 
and $110.0^\circ$ were used to determine the beam polarizations
for 135 and 70 MeV/u, respectively.
In Figs.~\ref{aij} and \ref{data_comp},
the new calibration data in Ref.~\cite{suda2002} 
are compared with the data in Refs.~\cite{sak96} 
and \cite{sek2002}. Only the statistical errors are shown.
These independent measurements, 
which used different methods to determine the
beam polarizations, 
provide a reasonably good agreement at 135 MeV/u.
However, there are systematic discrepancies at 70 MeV/u.
This disagreement is due to the systematic uncertainties 
for determining the polarization axis (less than $5\%$) 
and the uncertainties in the magnitudes of the beam polarizations 
(less than $4\%$).
The re-analyzed data at 70 MeV/u, which were obtained using the 
new calibration data are tabulated 
in Tables~\ref{tab140_d} and \ref{tab140_s}
and shown in Fig.~\ref{data_comp3} with the statistical errors.
Open diamonds (open triangles) in Fig.~\ref{data_comp3}
are the results measured with the SMART system (D-room polarimeter).
The newly analyzed data are in a reasonable agreement  
with the calibration data in Ref.~\cite{suda2002}.
It should be noted that the $A_{yy}$ 
at $\theta_{\rm c.m.} = 116.9^\circ$ of 135 MeV/u
was reanalyzed in a similar way using deuteron 
beam polarizations measured at 
$\theta_{\rm c.m.} = 86.5^\circ$ for $d$--$p$ elastic scattering.
These new polarizations reduced $A_{yy}$ at $\theta_{\rm c.m.} =
116.9^\circ$, which is shown with a open square in Fig.~\ref{aij}, 
by about 9\%.

\section{Theoretical formalism and dynamical input}

In this paper we study elastic $Nd$
scattering with the initial state $\phi$
composed of a deuteron 
and a nucleon.
The outgoing state $\phi'$ corresponds to a change of the outgoing nucleon
momentum. 
Using the matrix element of
the elastic scattering transition operator $U$ which is defined as
\begin{equation}
\langle \phi' \vert U \vert \phi \rangle \ = \ \langle \phi' \vert
P G_0^{-1} \ + \ V_4^{(1)} \, ( 1 + P ) \ + \ 
\ P T \  + \
V_4^{(1)} \, ( 1 + P ) \, G_0 \, T \vert \phi \rangle ,
\label{eqU1}
\end{equation}
the various spin observables and differential cross section can be
calculated~\cite{ohl72,glo96}. The quantity $G_0$ is the free 3N propagator
and $P$ takes into account the identity of nucleons and is the sum of a
cyclical and an anticyclical permutation of three nucleons. 
$V_4^{(1)}$ represents one of the terms of 
the 3N force $V_4$ 
\begin{equation}
\label{eqU2}
V_4 = V_4^{(1)} + V_4^{(2)} + V_4^{(3)} ,
\end{equation}
where each $V_4^{(i)}$ is symmetric under the exchange of the nucleons
$jk$ with $j\ne i \ne k$. In the $2\pi$-exchange 3NF, $V_4^{(1)}$ is
a contribution to the 3N potential from (off-shell) rescattering
of a pion on nucleon 1.
The first term in Eq.~(\ref{eqU1}) is a single nucleon exchange contribution
and is followed by a single interaction of three nucleons via
the 3NF. The remaining part results from rescattering
among three nucleons induced by two- and three-nucleon forces.
All these rescatterings are summed up in the integral equation
for the amplitude $T$~\cite{glo96,hub97}
\begin{equation}
\label{eqU3}
T \ = \ t \, P \, \phi \ 
+ \ ( 1 + t G_0 ) \, V_4^{(1)} \, ( 1 + P ) \, \phi \
+  \ t \, P \, G_0 \, T \
+ \ ( 1 + t G_0 ) \, V_4^{(1)} \, ( 1 + P ) \, G_0 \, T ,
\end{equation}
where the NN $t$-operator is denoted by $t$.
After projecting on a partial-wave momentum-space basis
this equation leads to a system of coupled integral equations which can
be solved numerically exactly for any nuclear force. In this study
we restricted our partial wave basis taking all states with the total angular
momenta $j$ in the two-nucleon subsystem smaller than 6. 
This
corresponds to a maximal number of 142 partial wave states in the
3N system for each total angular momentum. For the
energies of the present paper this provides 
 convergent results for the elastic
scattering observables. We checked that the convergence has been
achieved by looking at the results obtained when $j=6$ states have
been included. This increases the number of states to 194. This
convergence check was done without 3NF. The inclusion of the 3NF
has been carried through for all total angular momenta of the 3N
system up to $J=13/2$ while the longer ranged 2N interactions
require states up to $J=25/2$. For the details of the formalism
and the numerical performance we refer to Refs.~\cite{glo96,wit88,hub93}.

In this study predictions of different nuclear force
models are shown. They consist of one of the NN forces: AV18, CDBonn, 
Nijmegen I, II and 93, and a 3NF.
Each of the NN interactions was combined 
with the $2\pi$-exchange TM 3NF model~\cite{tm3nf}.
The combinations use the cut-off parameter $\Lambda$ in 
the strong form factor parameterization separately
adjusted to the $\rm ^3H$
binding energy for the different NN forces~\cite{nogga97}.
The $\Lambda$-values used with the AV18,
CDBonn, Nijmegen I, II, and 93 potentials are in that order 
$\Lambda =  5.215$, $4.856$, $5.120$, $5.072$, and $5.212$\@ 
(in units of $m_{\pi}$), respectively. 
The standard parameterization of the TM 3NF is
criticized in Refs.~\cite{fri99,huba1,yang74} since 
it violates chiral symmetry. 
A form more consistent with chiral 
symmetry was proposed by modifying the $c$-term of the TM force and 
absorbing the long range part of this term  into the $a$-term and 
rejecting the rest of the $c$-term~\cite{fri99,huba1}. 
This new form is called TM$'$(99)~\cite{tm99}.
The $\Lambda$-values used (again in units of $m_{\pi}$)
with the AV18,
CDBonn, Nijmegen I, and II potentials are $\Lambda =
4.764$, $4.469$, $4.690$, and $4.704$, respectively. 

For the AV18 potential we also use  the Urbana IX 3NF~\cite{urix}.
That force is based on the Fujita-Miyazawa assumption of an
intermediate $\Delta$ excitation in the $2\pi$
exchange~\cite{fuj57}, which is augmented by a phenomenological
spin-independent short-range part. This force is formulated in
configuration space~\cite{urix}. 
Refer to Ref.~\cite{wit01}  
for the partial wave
decomposition of the Urbana IX 3NF in momentum space.

\section{Discussion}

\subsection{Comparison of present data with theoretical predictions}

In Figs.~\ref{kij} and \ref{kij2} the theoretical predictions for
the five different NN potentials and their combinations with the
chosen 3NF's are shown for the polarization transfer
coefficients and the induced polarization. The light shaded bands
in Fig.~\ref{kij} are the results of the Faddeev calculations
based on the high precision NN potentials,
AV18, CDBonn,  Nijmegen I, II and 93 only. 
The dark shaded bands in Fig.~\ref{kij} contain
the predictions of the four NN forces with the TM$'$(99) 3NF. 
In each case the triton binding energy was adjusted to the
experimental value.  The solid lines in Fig.~\ref{kij} are the
theoretical predictions obtained using the AV18 potential combined
with the Urbana IX 3NF\@.
To avoid making the figure too complicated
the predictions combining the five NN
forces with the TM 3NF are shown in Fig.~\ref{kij2}
together with the calculations of TM$'$(99) 3NF.
The TM 3NF predictions are shown as light 
shaded bands and the TM$'$(99) 3NF ones  
are shown as dark shaded bands in that figure.

At first, theoretical predictions are separately compared to the data for
the polarization transfer coefficients $K_{xx}^{y'}$ and $K_{yy}^{y'}$.
For $K_{xx}^{y'}$, the 3NF effects are rather modest 
and the differences among the various 3NFs are small.
However, the data apparently prefer the 3NF predictions rather 
than the pure 2N force ones.
The deviation of the 3NF predictions from those for the 2N forces
is clearly pronounced for $K_{yy}^{y'}$, and 
the 2N band significantly overestimates  
the data at the angles $\theta_{\rm c.m.} = 90^\circ$--$120^\circ$. 
The inclusion of the Urbana IX 3NF provides a good description of the data.
Also the TM$'$(99) does fairly well, whereas the TM provides a 
better description of the data.
One can see clearly the difference between the data and 
the 2N force predictions for the polarization transfer coefficient
$K_{xx}^{y'} - K_{yy}^{y'}$.
The 2N force predictions underestimate the data 
in the region of the $K_{xx}^{y'}-K_{yy}^{y'}$ minima at the angles
$\theta_{\rm c.m.} = 90^\circ - 120^\circ$.
The inclusion of the Urbana IX 3NF as well as TM$'$(99) 
removes these discrepancies. Also for the TM 3NF
there is a good agreement between the data and theory.
For $K_{y}^{y'}$, at backward angles $\theta_{\rm c.m.}\ge 150^\circ$
the data support the NN forces only predictions as well as 
the TM$'$(99) and Urbana IX 3NFs ones.
In the angular range of $\theta_{\rm c.m.} = 90^\circ - 120^\circ$,
a large discrepancy exists between  the 2N force only predictions
and data.
The inclusion of either TM$'$(99) or Urbana IX 3NF shifts 
the calculated results 
in the right direction, but not enough to
describe the data. 
The effects  of the TM 3NF are also 
not sufficient to provide a good description of the data.
For $K_{xz}^{y'}$, the situation is complicated 
throughout the entire measured angular range 
and the data are not described by the theoretical predictions.
At backward angles $\theta_{\rm c.m.} \ge 150^\circ$,
the 2N band provides a moderate agreement to the data.
It clearly deviates from the data in the minimum region 
around $\theta_{\rm c.m.} = 100^\circ$,
but the predictions from the 3NFs used do not explain this discrepancy.
It is interesting to note that all 3NF models studied 
predict large effects in the region of this minimum, however the effects 
of the TM$'$(99) and Urbana IX 3NFs are in the opposite 
direction to those of the TM 3NF.
For the induced polarization $P^{y'}$
the 2N band overestimates the data around the region of 
the $P^{y'}$ maximum. The inclusion of the Urbana IX or 
TM$'$(99) 3NF brings the predictions closer to the data, 
while the TM 3NF provides large, incorrect effects.
For the analyzing powers in the $p$--$d$ elastic scattering
at incoming nucleon energies larger than about $60~\rm MeV$,
a similar pattern of  discrepancies 
between the data and theoretical predictions 
is found~\cite{erm01,wit01}.

The predictions including the TM$'$(99) 3NF, 
which were not presented in our previous study \cite{sek2002}, 
are compared with our deuteron analyzing powers at 135 and 70 MeV/u
in Figs.~\ref{aij} and~\ref{data_comp3}, respectively.
Comparison of the vector and tensor analyzing powers to the 
TM$'$(99) predictions shown in Figs.~\ref{aij} and \ref{data_comp3}
reveals that the effects of the TM$'$(99) 3NF are similar 
in size and directions to the effects of the Urbana IX 3NF,
except for $A_{xz}$ at $135~\rm MeV/u$.

\subsection{Summary of the comparison between 
{\mbox{\boldmath $d$--$p$}} polarization data and Theoretical Predictions}

In this section we would like to summarize the comparison of the
theoretical predictions to $d$--$p$ elastic scattering data
reported here and in Ref.~\cite{sek2002}. It encompasses all
deuteron analyzing powers $A_y^d$, $A_{xx}$, $A_{yy}$, $A_{xz}$,
the proton induced polarization $P^{y'}$ $\left( = -
A_y^{p}\right)$, and the deuteron to proton polarization transfer
coefficients $K_y^{y'}$, $K_{xx}^{y'} - K_{yy}^{y'}$
($K_{xx}^{y'}$, $K_{yy}^{y'}$), and $K_{xz}^{y'}$.

Generally, the discrepancies between the data and the pure 2N
force predictions are clearly seen at the angles
where the cross sections have minima. For the cross sections
these discrepancies at the two energies considered here are
explained by taking into account the 2$\pi$ exchange type 3NF
models (TM, TM$'$(99), and Urbana IX ). Thus all $2\pi$-exchange 3NF
potentials considered here (TM, TM$'$(99), and Urbana IX ) provide
3NF effects for the cross sections which are comparable in
magnitude and sign. At higher energies, however, discrepancies
remain to the data in the minima and even more at backward
angles Ref.~\cite{hatanaka02,erm2003}.

Spin observables can be grouped into three types.
The {\bf Type I} observables are the deuteron
vector analyzing power $A_y^{d}$ and the deuteron to proton
polarization transfer coefficient  $K_{xx}^{y'} - K_{yy}^{y'}$
($K_{xx}^{y'}$, $K_{yy}^{y'}$).
The deviations between the data and the 2N force predictions for
these observables are explained by the inclusion of the
2$\pi$-exchange 3NFs considered here (TM, TM$'$(99), Urbana IX )
similarly as in the case of the cross section. These observables provide 
a clear evidence for the 3NFs.

The {\bf Type II} observable is the proton induced polarization $P^{y'}$,
which is equivalent to the proton analyzing power $A_y^{p}$
$\left( P^{y'}= - A_y^{p} \right)$.
The TM$'$(99) 3NF and Urbana IX 3NF describe
the difference between the data and the 2N force predictions.
The inclusion of the TM 3NF shifts the calculated
results in the right direction, but the effects are too large.
The nonzero $c$--term of the TM 3NF
might be the origin of the incorrect 3NF effect.
In order to see this more clearly, it is interesting to identify
the effects due to the intermediate $\Delta$--isobar excitation which is
the main part of the 2$\pi$--exchange 3NF. Recently, the Hannover
group carried out calculations which explicitly included
the $\Delta$--isobar excitation in the framework of the coupled channel
approach~\cite{arnas2003}. In their calculations, the CDBonn
potential was taken as the 2N interaction.
One can get directly the $\Delta$-isobar effects (magnitude and/or
direction) by comparing their predictions with and without the
$\Delta$-isobar excitation. The results are shown in
Figs.~\ref{kij_hv} and \ref{aij_hv}. 
The $\Delta$-isobar effects are similar to those of the TM$'$(99) and
Urbana IX 3NFs for almost all observables except for $A_{xx}$.
This feature indicates that the poor agreement for TM in $P^{y'}$
of the {\bf Type II} is not due to the $2\pi$-exchange
$\Delta$-isobar excitation. Since the main difference between the
TM and TM$'$(99) 3NFs comes from the non-vanishing  $c$-term in the TM
3NF, this term is most probably responsible for the poor description  of
$P^{y'}$ by the TM 3NF.

The {\bf Type III} observables are the deuteron tensor analyzing
powers $A_{xx}$, $A_{yy}$, $A_{xz}$, and the deuteron to proton
polarization transfer coefficients $K_{y}^{y'}$, and
$K_{xz}^{y'}$. No calculation shows a superiority for these
observables.
       Although large effects of 3NFs are predicted at
       the angles $\theta_{\rm c.m.} = 90^\circ - 120^\circ$,
       they are not supported by the data.
       It is interesting to note that the TM$'$(99) and
       Urbana IX 3NFs provide very similar effects.
       On the other hand, the effects induced by the TM 3NF
       are quite different from the TM$'$(99) and Urbana IX 3NF's ones.
The {\bf Type III} observables clearly reveal the defects of the
present day 3NF models. To describe these spin observables one
should look for other 3NF terms in addition to the $2\pi$-exchange
3NFs. At low energies, Wita{\l}a {\it et al.} applied 3NFs
based on $\pi$--$\rho$ and $\rho$--$\rho$ exchanges
\cite{tm3nf2,ref28} and found their effects on cross sections and
spin observables for $Nd$ elastic scattering~\cite{wit95}. It was
found that the effects of the $\pi - \rho$ exchange generally 
reduced the effects caused by the $2\pi$-exchange TM
3NF. The effects induced by $\rho - \rho$ exchanges were
negligible. It would be interesting to apply these $\pi$--$\rho$
and $\rho$--$\rho$ exchange 3NFs also at intermediate energies
where the interferences might be different.
Recently  new 3NF models called the Illinois model have been
reported~\cite{piep01} and found to be successful in describing
the binding and excitation energies of light nuclei with mass
number up to $\rm A = 10$. This model is an extension of the
Urbana IX 3NF and consists of five terms: the two-pion-exchange
terms due to $\pi N$ scattering in $S$ and $P$ waves, a
phenomenological repulsive term, and the three-pion-exchange terms
($V^{3\pi,\Delta R}$) due to ring diagrams with $\Delta$ in the
intermediate states. The $V^{3\pi,\Delta R}$ is a new type of 3NF
and contains new spin dependent terms, such as 
$\vec{\sigma}\cdot(\vec{{\textit{\textbf r}}}_{ij}
     \times \vec{{\textit{\textbf r}}}_{jk} )$ term.
These spin dependent terms might explain {\bf
Type III} spin observables. It would be interesting to include
these new terms into the 3N continuum calculations.
The results of the coupled channel formulations with the
$\Delta$-isobar excitations are supported by the tensor analyzing
power $A_{xx}$~\cite{arnas2003}, which TM$'$(99) as well as Urbana IX
3NFs do not describe well. This points to contributions which are
not included in 2$\pi$-exchange 3NF models.

As the incident nucleon energy increases one should not ignore
relativity which becomes  more and more important. Some
indications on their importance was found in the analysis of the
high precision $nd$ total cross section data~\cite{wit99a} and in
the study of backward angular distributions of $p$--$d$ elastic
scattering cross section at higher energies of the incident
nucleon \cite{hatanaka02,erm2003}. The discrepancies between the data and
nonrelativistic predictions become larger with increasing energy
and cannot be removed by including different 3NFs~\cite{wit99a}. 
Therefore relativity might be another candidate to provide 
a solution for the {\bf Type III} spin observables and 
such treatment is now proceeding  \cite{kam98,kam2002}.

In chiral perturbation theory at
NNLO~\cite{epel2002} the $2\pi$-exchange 3NF 
together with a one-pion exchange between a NN contact force and 
the third nucleon and a pure 3N contact force occurs. 
This also suggests that the $2\pi$-exchange 
should be supplemented by the exchange of a pion
together with heavy mesons and the exchange of two heavy mesons.
In addition quite a few types of 3NFs appear at NNNLO, which have
to be worked out.
This will lead  to additional spin-dependences, which 
will be required for the {\bf Type III} observables.

\section{Summary and Conclusion}

The 
deuteron to proton ($\vec{d} + p \to \vec{p} + d$) 
polarization transfer coefficients
$K_{y}^{y'}$, $K_{xx}^{y'}-K_{yy}^{y'}$ ($K_{xx}^{y'}$, $K_{yy}^{y'}$), 
and $K_{xz}^{y'}$ were measured at 135 MeV/u
in the angular range $\theta_{\rm c.m.} = 90^\circ$--$180^\circ$.
The induced proton polarization $P^{y'}$ was also measured.
The statistical uncertainties are smaller than $0.03$
for all the polarization transfer coefficients,
and $0.02$ for the induced polarization $P^{y'}$.
The estimated systematic uncertainties for the polarization 
transfer coefficients and the induced polarization $P^{y'}$
are about 8\%.
The induced polarization $P^{y'}$ was 
compared with the analyzing power $A_y^{p}$ 
for the time reversed reaction, 
$^2{\rm H}(\vec{p},p)^2{\rm H}$ elastic scattering, measured at KVI.
The data are consistent within 
the statistical uncertainties in the measured angular range.

In our previous study,
the measurements of a complete set of deuteron analyzing powers %
were measured 
at incoming deuteron energies 70, 100, and 135~MeV/u,
covering a wide angular range $\theta_{\rm c.m.} = 10^\circ$--$180^\circ$.
Also the unpolarized cross sections were   %
measured at the same angles  at 70 and 135 MeV/u.
High precision data have been obtained.

Our data are compared with predictions based 
on different modern nuclear forces in order to look 
for evidence of 3NF effects and 
to test  present day 3NF models with                                         %
respect to the effects that these forces cause. 
Based on the comparison of our data with pure 2N force predictions
clear discrepancies, which increase with deuteron energy, 
are found for most observables, 
especially at the angles around the cross section minimum.
Including any of the $2\pi$-exchange 3NFs used in the present paper,
the TM 3NF, the modified version of it TM$'$(99), and the Urbana IX 3NF,
can reduce the discrepancies observed for the cross section, 
for the deuteron vector analyzing power $A_y^{d}$,
and for the polarization transfer coefficient 
$K_{xx}^{y'} - K_{yy}^{y'}$ ($K_{xx}^{y'}$, $K_{yy}^{y'}$).
Thus, these observables can be considered to provide 
a clear evidence for the 3NF effects.
For the induced polarization $P^{y'}$,
the TM$'$(99) and Urbana IX 3NFs explain the difference between 
the data and the 2N force predictions. On the other hand, 
the TM 3NF fails to describe this observable.
This appears to indicate that
the non-vanishing $c$--term of the TM 3NF, which should not exist
according to chiral symmetry,
is probably responsible for the failure of the model.
For the tensor analyzing powers and the polarization 
transfer coefficients $K_y^{y'}$ and $K_{xz}^{y'}$,
calculations unsuccessfully describe the data.
Large effects of 3NFs are predicted at the angles
$\theta_{\rm c.m.} = 90^\circ - 120^\circ$. 
However, the data do not support the these predictions.
Our results clearly reveal the defects of the present day 3NFs. 

Finally it should be noted that 
this is the first precise data set of the 
analyzing powers, 
and polarization transfer coefficients
for $d$--$p$ elastic scattering at intermediate energies,
which will provide a solid basis to test  future
3NF models.

\begin{acknowledgments}
We acknowledge the outstanding work of the RIKEN Accelerator group
for delivering excellent polarized deuteron beams. 
We thank P.\ U.\ Sauer and A.\ Deltuva for providing their results
as shown in this paper and for their useful comments.
K.\ S.\ would like to acknowledge the support of 
Special Postdoctoral Researchers Program of RIKEN\@. 
N.\ K. and K.\ E.\ would like to thank RIKEN for the hospitality 
during their stay in Japan.
This work was supported financially in part by the Grants-in-Aid 
for Scientific Research Numbers 04402004 and 10304018 of the Ministry of
Education, Culture, Sports, Science, and Technology of Japan, 
the Polish Committee for Scientific Research under
Grant No.\ 2P03B00825, and DOE grant No's DE-FG03-00ER41132 and 
DE-FC02-01ER41187.
The numerical calculations were performed on the 
SV1 and the CRAY T3E of the NIC in J\"ulich, Germany.  
\end{acknowledgments}

\newpage 
\bibliography{apssamp}

\subsection*{FIGURES}
\subsection*{TABLES}

\begingroup
\squeezetable
\endgroup

\begin{figure}[htbp]
\caption{Arrangement of the RIKEN Spectrograph SMART.
The FP1 and FP2 denote the first and 
second focal planes, respectively.
Scattered protons were momentum analyzed by the magnetic
spectrograph and detected at the FP2. 
The polarizations of the scattered protons were
measured with the focal plane polarimeter DPOL.
\label{smart}
}
\end{figure}

\begin{figure}[htbp]
\caption{Second-focal-plane detector system including 
the focal plane polarimeter DPOL. It consists of
two multiwire drift chambers 
(\@MWDC\@1 and MWDC\@2 ), 
the plastic scintillation trigger counters ( SC1, SC2, SC3 ), 
a polarization analyzer target, and the counter hodoscope system 
( HOD and CM ).
\label{dpol}
}
\end{figure}

\begin{figure}[htbp]
\caption{Definition of the Coordinate Frame for 
the SMART system. 
The $\{p_{ij}\}$ denotes the vector or tensor deuteron beam polarizations.
The $p_{i'}$ is the polarization of the scattered protons and 
the $p_{i''}$ is the polarization of 
the scattered protons at the second focal plane FP2\@.
\label{smartconfig}}
\end{figure}

\begin{figure}[htbp]
\caption{Energy dependence of the effective analyzing power
$A_y^C$ of the DPOL and the measured data.
\label{aycfit}}
\end{figure}

\begin{figure}[htbp]
\caption{Excitation energy spectra for $d$--$p$ elastic scattering 
at c.m. angles $\theta_{\rm c.m.} = 176.8^\circ,~120^\circ,{\rm and}~90^\circ$
taken with the liquid hydrogen target.}
\label{ex_LiQ}
\end{figure}

\begin{figure}[htbp]
\caption{(Color online) polarization transfer coefficients $K_{xx}^{y'}$,
$K_{yy}^{y'}$, $K_{xx}^{y'} - K_{yy}^{y'}$,  $K_{y}^{y'}$,
$K_{xz}^{y'}$, and the induced polarization $P^{y'}$ in elastic
$d$--$p$ scattering at the incident deuteron energy of 135 MeV/u.
Open circles are the data in the present measurement and
open squares are the data in the test measurement~\cite{sak2000}.
Solid squares on the $P^{y'}$ figure  are the proton analyzing
power data for the time-reversed reaction 
$^2{\rm H}(\vec{p},p)^2{\rm H}$~\cite{erm01}. 
The light shaded bands contain the NN force
predictions (AV18, CDBonn, Nijmegen I, II and 93), and the dark shaded
bands contain the combinations of the NN + TM$'$(99) 3NF predictions as
described in the text. The solid line is the AV18 + Urbana IX 3NF
prediction. \label{kij} 
}
\end{figure}

\begin{figure}[htbp]
\caption{(Color online) polarization transfer coefficients $K_{xx}^{y'}$,
$K_{yy}^{y'}$, $K_{xx}^{y'} - K_{yy}^{y'}$,  $K_{y}^{y'}$,
$K_{xz}^{y'}$, and the induced polarization $P^{y'}$ in elastic
$d$--$p$ scattering at 135 MeV/u.
The light shaded bands contain the combinations of the NN +
TM force predictions  while the dark shaded bands include the
combinations with TM$'$(99). For the descriptions of 
symbols, see Fig.~\ref{kij}.
\label{kij2} }
\end{figure}

\begin{figure}[htbp]
\caption{(Color online) vector and tensor deuteron analyzing powers for $d$--$p$
 elastic scattering at 135 MeV/u reported in \cite{sek2002,suda2002}.
Solid circles are the new calibration data of \cite{suda2002} while
open squares and circles are the results presented in
\cite{sek2002}. For the descriptions of bands and curves, see
Fig.~\ref{kij}. \label{aij} 
}
\end{figure}

\begin{figure}[htbp]
\caption{(Color online) deuteron analyzing powers at 70 MeV/u.
For descriptions of symbols, see Fig.~\ref{aij}.
\label{data_comp}}
\end{figure}

\begin{figure}[htbp]
\caption{(Color online) deuteron analyzing powers at 70 MeV/u
obtained using the new calibration data.
Open diamonds are the results measured with the SMART system
and open triangles show the results measured with the D-room polarimeter.
For the descriptions of bands and curves, see Fig.~\ref{kij}.
\label{data_comp3}}
\end{figure}

\begin{figure}[htbp]
\caption{Polarization transfer coefficients $K_{xx}^{y'}$,
$K_{yy}^{y'}$, $K_{xx}^{y'} - K_{yy}^{y'}$, $K_{y}^{y'}$,
$K_{xz}^{y'}$, and the induced polarization $P^{y'}$ in elastic
$d$--$p$ scattering at 135 MeV/u.
The solid lines are the coupled channel approach predictions
obtained with $\Delta$-isobar excitations and the dotted lines
based on the CDBonn potential~\cite{arnas2003}. 
For the descriptions of symbols, see Fig.~\ref{kij}.
\label{kij_hv}
}
\end{figure}

\begin{figure}[htbp]
\caption{Vector and tensor deuteron analyzing powers
in elastic $d$--$p$ scattering at 135 MeV/u.
For the descriptions of lines see Fig.~\ref{kij_hv}.
For the descriptions of symbols, see Fig.~\ref{aij}.
\label{aij_hv}
}
\end{figure}

\begin{table}[htbp]
\centering
 \begin{tabular}{rrcrcrcrcrc} \hline \hline
  \multicolumn{1}{c}{$\theta_{\rm c.m.}\rm [deg]$}  
  &\multicolumn{1}{c}{$~P^{y'}~$}      & \multicolumn{1}{c}{$~\delta P^{y'}~$}                  
  &\multicolumn{1}{c}{$~~K_{y}^{y'}~~$}  & \multicolumn{1}{c}{$~\delta K_{y}^{y'}~$} 
  &\multicolumn{1}{c}{$~K_{yy}^{y'}~$} & \multicolumn{1}{c}{$~\delta K_{yy}^{y'}~$} 
  &\multicolumn{1}{c}{$~K_{xx}^{y'}~$} & \multicolumn{1}{c}{$~\delta K_{xx}^{y'}~$} 
  &\multicolumn{1}{c}{$~K_{xz}^{y'}~$} & \multicolumn{1}{c}{$~\delta K_{xz}^{y'}~$} 
\\ \hline 
 90.0&    0.495 &   0.010 &   0.162 &   0.017 &   0.385 &   0.026 &  $-$0.492 &   0.031 &
  $-$0.286 &   0.018 \\
100.0&    0.532 &   0.013 &   0.256 &   0.014 &   0.454 &   0.020 &  $-$0.348 &   0.020 &
  $-$0.310 &   0.023 \\
110.0&    0.481 &   0.015 &   0.306 &   0.015 &   0.482 &   0.022 &  $-$0.231 &   0.023 &
  $-$0.290 &   0.019 \\
120.0&    0.347 &   0.008 &   0.323 &   0.013 &   0.416 &   0.016 &   0.096 &   0.027 &
  $-$0.245 &   0.021 \\
130.0&    0.116 &   0.008 &   0.229 &   0.027 &   0.230 &   0.028 &   0.267 &   0.027 &
   0.080 &   0.020 \\
140.0&   $-$0.080 &   0.009 &   0.111 &   0.013 &  $-$0.115 &   0.015 &   0.356 &   0.029 &
   0.338 &   0.022 \\
150.0&   $-$0.214 &   0.010 &   0.156 &   0.012 &  $-$0.359 &   0.015 &   0.266 &   0.015 &
   0.358 &   0.021 \\
160.0&   $-$0.185 &   0.006 &   0.434 &   0.019 &  $-$0.215 &   0.024 &   0.073 &   0.015 &
   0.122 &   0.023 \\
170.0&   $-$0.089 &   0.006 &   0.654 &   0.019 &  $-$0.065 &   0.021 &   0.017 &   0.020 &
  $-$0.039 &   0.022 \\
176.8&   $-$0.014 &   0.005 &   0.687 &   0.014 &  $-$0.022 &   0.015 &   0.023 &   0.021 &
  $-$0.057 &   0.021 \\
\hline \hline
\end{tabular}
\caption{Data table for $d$--$p$ elastic scattering
 deuteron to proton polarization transfer
 coefficients  and induced proton polarizations at $135~\rm MeV/u$.\label{kij_tab}}
\end{table}

\begin{table}[htbp]
\centering
 \begin{tabular}{rrccrccrccrcc} \hline \hline
$\theta_{\rm c.m.}$ & 
\multicolumn{1}{c}{$~A_{y}^d$} & 
\multicolumn{1}{c}{$~\delta A_{y}^{d(\rm st)}$} & 
\multicolumn{1}{c}{$\delta A_{y}^{d(\rm sy)}$} &
\multicolumn{1}{c}{$~~A_{yy}$} & 
\multicolumn{1}{c}{$~\delta A_{yy}^{(\rm st)}$} & 
\multicolumn{1}{c}{$~\delta A_{yy}^{(\rm sy)}$} &
\multicolumn{1}{c}{$~A_{xx}$} & 
\multicolumn{1}{c}{$~\delta A_{xx}^{(\rm st)}$}& 
\multicolumn{1}{c}{$~\delta A_{xx}^{(\rm sy)}$} &
\multicolumn{1}{c}{$~~A_{xz}$} & 
\multicolumn{1}{c}{$~\delta A_{xz}^{(\rm st)}$} & 
\multicolumn{1}{c}{$~\delta A_{xz}^{(\rm sy)}$} 
\\ \hline
$82.0$ &
  $-0.309$ &  $~0.001$ &  $0.008$ &
  $~0.246$ &  $~0.001$ &  $0.006$ & 
  $-0.225$ &  $~0.001$ &  $0.006$ & 
  $~0.253$ &  $~0.013$ &  $0.029$ \\ 
$88.0$ &
  $-0.403$ &  $~0.001$ &  $0.010$ &
  $~0.312$ &  $~0.001$ &  $0.008$ & 
  $-0.207$ &  $~0.001$ &  $0.005$ & 
  $~0.320$ &  $~0.014$ &  $0.037$ \\ 
$94.0$ &
  $-0.477$ &  $~0.001$ &  $0.012$ & 
  $~0.383$ &  $~0.001$ &  $0.009$ & 
  $-0.168$ &  $~0.001$ &  $0.004$ & 
  $~0.377$ &  $~0.015$ &  $0.044$ \\ 
$100.0$ &
  $-0.514$ &  $~0.001$ &  $0.013$ &
  $~0.452$ &  $~0.001$ &  $0.011$ & 
  $-0.111$ &  $~0.001$ &  $0.003$ & 
  $~0.407$ &  $~0.017$ &  $0.047$ \\ 
$110.0$ &
  $-0.488$ &  $~0.002$ &  $0.012$ &
  $~0.542$ &  $~0.002$ &  $0.013$ & 
  $-0.001$ &  $~0.002$ &  $0.001$ & 
  $~0.366$ &  $~0.026$ &  $0.043$ \\ 
$119.1$ &
  $-0.383$ &  $~0.002$ &  $0.011$ & 
  $~0.578$ &  $~0.002$ &  $0.017$ & 
  $~0.067$ &  $~0.002$ &  $0.004$ & 
  $~0.213$ &  $~0.028$ &  $0.047$ \\
\hline \hline
\end{tabular}
\caption{Calibrated analyzing power data for
 $d$--$p$ elastic scattering at $70~\rm MeV/u$ 
 reported in Ref.~\cite{suda2002}.
 The $\delta A_{ij}^{(\rm st)}$ denotes the statistical
 error and the $\delta A_{ij}^{(\rm sy)}$ denotes the
 systematic one.
 \label{aij_suda70}
 }
\end{table}

\begin{table}[htbp]
\centering
 \begin{tabular}{rrccrccrccrcc} \hline \hline
$\theta_{\rm c.m.}$ & 
\multicolumn{1}{c}{$~A_{y}^d$} & 
\multicolumn{1}{c}{$~\delta A_{y}^{d(\rm st)}$} & 
\multicolumn{1}{c}{$\delta A_{y}^{d(\rm sy)}$} &
\multicolumn{1}{c}{$~~A_{yy}$} & 
\multicolumn{1}{c}{$~\delta A_{yy}^{(\rm st)}$} & 
\multicolumn{1}{c}{$~\delta A_{yy}^{(\rm sy)}$} &
\multicolumn{1}{c}{$~A_{xx}$} & 
\multicolumn{1}{c}{$~\delta A_{xx}^{(\rm st)}$}& 
\multicolumn{1}{c}{$~\delta A_{xx}^{(\rm sy)}$} &
\multicolumn{1}{c}{$~~A_{xz}$} & 
\multicolumn{1}{c}{$~\delta A_{xz}^{(\rm st)}$} & 
\multicolumn{1}{c}{$~\delta A_{xz}^{(\rm sy)}$} 
\\ \hline
$80.6$ &
   $-0.345$ & $~0.001$ & $0.011$ & 
   $~0.398$ & $~0.001$ & $0.012$ & 
   $-0.494$ & $~0.001$ & $0.015$ & 
   $~0.405$ & $~0.005$ & $0.036$ \\
$83.6$ &
  $-0.374$ &  $~0.001$ & $0.011$ & 
  $~0.424$ &  $~0.001$ & $0.013$ & 
  $-0.481$ &  $~0.001$ & $0.014$ & 
  $~0.433$ &  $~0.005$ & $0.039$ \\ 
$86.6$ &
  $-0.393$ &  $~0.001$ &  $0.012$ & 
  $~0.446$ &  $~0.001$ &  $0.013$ & 
  $-0.471$ &  $~0.001$ &  $0.014$ & 
  $~0.449$ &  $~0.005$ &  $0.040$ \\ 
$89.6$ &
 $-0.413$ &  $~0.001$ &  $0.013$ &
 $~0.469$ &  $~0.001$ &  $0.014$ & 
 $-0.457$ &  $~0.001$ &  $0.014$ & 
 $~0.454$ &  $~0.005$ &  $0.041$ \\ 
$92.6$ &
  $-0.420$ & $~0.001$ &  $0.013$ &
  $~0.498$ & $~0.001$ &  $0.015$ & 
  $-0.442$ & $~0.001$ &  $0.013$ & 
  $~0.460$ & $~0.006$ &  $0.041$ \\ 
$117.7$ &
  $-0.346$ & $~0.002$ &  $0.012$ &
  $~0.628$ & $~0.002$ &  $0.018$ & 
  $-0.327$ & $~0.002$ &  $0.010$ & 
  $~0.478$ & $~0.008$ &  $0.043$ \\ 
\hline \hline
\end{tabular}
\caption{Calibrated analyzing power data
 for $d$--$p$ elastic scattering at $135~\rm MeV/u$
 reported in Ref.~\cite{suda2002}. 
 For the descriptions of the $\delta A_{ij}^{(\rm st)}$
 and $\delta A_{ij}^{(\rm sy)}$, see Table~\ref{aij_suda70}.
\label{aij_suda135}}
\end{table}

\begin{table}[htbp]
\centering
 \begin{tabular}{rrcrcrcrc} \hline \hline
  \multicolumn{1}{c}{$\theta_{\rm c.m.}\rm [deg]$}  
  &\multicolumn{1}{c}{$~A_y^d~$}      & \multicolumn{1}{c}{$\delta A_y^d$}                  
  &\multicolumn{1}{c}{$~A_{yy}~$}      & \multicolumn{1}{c}{$\delta A_{yy}$}                  
  &\multicolumn{1}{c}{$~A_{xx}~$}      & \multicolumn{1}{c}{$\delta A_{xx}$}                  
  &\multicolumn{1}{c}{$~A_{xz}~$}      & \multicolumn{1}{c}{$\delta A_{xz}$} 
\\ \hline 
$ 65.0$ & $      -0.016$ & $       ~0.002$ & $       ~0.121$ & $       ~0.004$ & $
          $ & $        $ & $      $ & $            $ \\
$ 70.1$ & $      -0.097$ & $       ~0.003$ & $       ~0.148$ & $       ~0.006$ & $
      -0.216$ & $       ~0.005$ & $       ~0.207$ & $       ~0.006$ \\
$ 75.0$ & $      -0.190$ & $       ~0.004$ & $       ~0.186$ & $       ~0.007$ & $
      -0.241$ & $       ~0.005$ & $       ~0.266$ & $       ~0.005$ \\
$ 80.0$ & $      -0.276$ & $       ~0.007$ & $       ~0.228$ & $       ~0.014$ & $
      -0.244$ & $       ~0.011$ & $       ~0.299$ & $       ~0.008$ \\
$ 85.0$ & $      -0.369$ & $       ~0.004$ & $       ~0.280$ & $       ~0.004$ & $
      -0.224$ & $       ~0.006$ & $       ~0.350$ & $       ~0.006$ \\
$ 88.2$ & $      -0.405$ & $       ~0.004$ & $       ~0.328$ & $       ~0.004$ & $
      -0.212$ & $       ~0.007$ & $       ~0.378$ & $       ~0.006$ \\
$ 90.0$ & $      -0.448$ & $       ~0.004$ & $       ~0.355$ & $       ~0.004$ & $
      -0.208$ & $       ~0.006$ & $      $ & $            $ \\
$ 95.0$ & $      -0.501$ & $       ~0.005$ & $       ~0.403$ & $       ~0.009$ & $
      -0.166$ & $       ~0.006$ & $       ~0.395$ & $       ~0.006$ \\
$100.0$ & $      -0.511$ & $       ~0.004$ & $       ~0.450$ & $       ~0.008$ & $
      -0.095$ & $       ~0.005$ & $       ~0.388$ & $       ~0.008$ \\
$105.0$ & $      -0.521$ & $       ~0.005$ & $       ~0.499$ & $       ~0.004$ & $
      -0.063$ & $       ~0.006$ & $       ~0.385$ & $       ~0.008$ \\
$110.0$ & $      -0.493$ & $       ~0.007$ & $       ~0.536$ & $       ~0.008$ & $
       0.015$ & $       ~0.005$ & $       ~0.368$ & $       ~0.012$ \\
$120.0$ & $      -0.352$ & $       ~0.006$ & $       ~0.577$ & $       ~0.012$ & $
       0.060$ & $       ~0.007$ & $       ~0.220$ & $       ~0.010$ \\
$130.0$ & $      -0.120$ & $       ~0.008$ & $       ~0.557$ & $       ~0.007$ & $
      -0.074$ & $       ~0.015$ & $       ~0.094$ & $       ~0.012$ \\
\hline \hline
\end{tabular}
\caption{Data table for 
 analyzing powers for $d$--$p$ elastic scattering at $70~\rm MeV/u$ 
 measured with the D-room polarimeter.
 \label{tab140_d}}
\end{table}

\begin{table}[htbp]
\centering
 \begin{tabular}{rrcrcrcrc} \hline \hline
  \multicolumn{1}{c}{$\theta_{\rm c.m.}\rm [deg]$}  
  &\multicolumn{1}{c}{$~A_y^d~$}      & \multicolumn{1}{c}{$\delta A_y^d$}                  
  &\multicolumn{1}{c}{$~A_{yy}~$}      & \multicolumn{1}{c}{$\delta A_{yy}$}                  
  &\multicolumn{1}{c}{$~A_{xx}~$}      & \multicolumn{1}{c}{$\delta A_{xx}$}                  
  &\multicolumn{1}{c}{$~A_{xz}~$}      & \multicolumn{1}{c}{$\delta A_{xz}$} 
\\ \hline 
$ 13.7$ & $       0.113$ & $       ~0.004$ & $       ~0.036$ & $       ~0.005$ & $
       0.039$ & $       ~0.004$ & $      $ & $            $ \\
$ 16.8$ & $       0.121$ & $       ~0.004$ & $       ~0.037$ & $       ~0.005$ & $
       0.016$ & $       ~0.005$ & $      $ & $            $ \\
$ 22.4$ & $       0.129$ & $       ~0.005$ & $       ~0.043$ & $       ~0.006$ & $
      -0.033$ & $       ~0.005$ & $      $ & $            $ \\
$ 28.6$ & $       0.163$ & $       ~0.005$ & $       ~0.055$ & $       ~0.006$ & $
      -0.056$ & $       ~0.005$ & $      $ & $            $ \\
$ 33.9$ & $       0.162$ & $       ~0.004$ & $       ~0.059$ & $       ~0.006$ & $
      -0.088$ & $       ~0.005$ & $       ~0.006$ & $       ~0.005$ \\
$ 37.1$ & $       0.171$ & $       ~0.005$ & $       ~0.068$ & $       ~0.006$ & $
      -0.103$ & $       ~0.005$ & $       ~0.008$ & $       ~0.006$ \\
$ 38.7$ & $       0.175$ & $       ~0.005$ & $       ~0.066$ & $       ~0.006$ & $
      -0.114$ & $       ~0.006$ & $       ~0.018$ & $       ~0.006$ \\
$ 40.3$ & $       0.169$ & $       ~0.005$ & $       ~0.070$ & $       ~0.007$ & $
      $ & $            $ & $      $ & $            $ \\
$ 41.9$ & $       0.173$ & $       ~0.006$ & $       ~0.071$ & $       ~0.007$ & $
      $ & $            $ & $      $ & $            $ \\
$ 44.5$ & $       0.165$ & $       ~0.003$ & $       ~0.077$ & $       ~0.004$ & $
      -0.111$ & $       ~0.005$ & $       ~0.029$ & $       ~0.007$ \\
$ 47.8$ & $       0.155$ & $       ~0.004$ & $       ~0.078$ & $       ~0.004$ & $
      -0.150$ & $       ~0.006$ & $      $ & $            $ \\
$ 51.2$ & $       0.135$ & $       ~0.004$ & $       ~0.089$ & $       ~0.004$ & $
      -0.167$ & $       ~0.009$ & $       ~0.061$ & $       ~0.007$ \\
$ 55.9$ & $       0.110$ & $       ~0.004$ & $       ~0.108$ & $       ~0.004$ & $
      -0.172$ & $       ~0.006$ & $       ~0.111$ & $       ~0.006$ \\
$ 59.4$ & $       0.072$ & $       ~0.004$ & $       ~0.112$ & $       ~0.004$ & $
      -0.204$ & $       ~0.005$ & $       ~0.134$ & $       ~0.005$ \\
$ 59.7$ & $       0.068$ & $       ~0.006$ & $       ~0.108$ & $       ~0.006$ & $
      $ & $            $ & $      $ & $            $ \\
$ 63.3$ & $      -0.001$ & $       ~0.006$ & $       ~0.106$ & $       ~0.006$ & $
      -0.214$ & $       ~0.008$ & $       ~0.160$ & $       ~0.011$ \\
$ 72.1$ & $      -0.134$ & $       ~0.005$ & $       ~0.163$ & $       ~0.005$ & $
      -0.221$ & $       ~0.009$ & $       ~0.216$ & $       ~0.013$ \\
\end{tabular}
\end{table}

\begin{table}
 \begin{tabular}{rrcrcrcrc} 
$121.8$ & $      -0.337$ & $       ~0.006$ & $       ~0.579$ & $       ~0.003$ & $
       0.077$ & $       ~0.006$ & $       ~0.196$ & $       0.011$ \\
$124.1$ & $      -0.330$ & $       ~0.011$ & $       ~0.578$ & $       ~0.006$ & $
       0.063$ & $       ~0.008$ & $       ~0.144$ & $       0.008$ \\
$126.1$ & $      -0.271$ & $       ~0.011$ & $       ~0.574$ & $       ~0.005$ & $
       0.048$ & $       ~0.008$ & $       ~0.112$ & $       0.013$ \\
$128.2$ & $      -0.192$ & $       ~0.011$ & $       ~0.572$ & $       ~0.008$ & $
      -0.003$ & $       ~0.008$ & $       ~0.093$ & $       0.012$ \\
$130.2$ & $      -0.150$ & $       ~0.011$ & $       ~0.559$ & $       ~0.008$ & $
      -0.035$ & $       ~0.008$ & $       ~0.081$ & $       0.012$ \\
$133.7$ & $      -0.040$ & $       ~0.007$ & $       ~0.547$ & $       ~0.006$ & $
      -0.138$ & $       ~0.006$ & $       ~0.037$ & $       0.005$ \\
$135.7$ & $       0.009$ & $       ~0.007$ & $       ~0.479$ & $       ~0.006$ & $
      -0.215$ & $       ~0.006$ & $       ~0.026$ & $       0.007$ \\
$138.4$ & $       0.097$ & $       ~0.006$ & $       ~0.448$ & $       ~0.005$ & $
      -0.334$ & $       ~0.011$ & $       ~0.026$ & $       0.006$ \\
$140.4$ & $       0.161$ & $       ~0.006$ & $       ~0.425$ & $       ~0.005$ & $
      -0.434$ & $       ~0.010$ & $       ~0.058$ & $       0.007$ \\
$142.5$ & $       0.181$ & $       ~0.005$ & $       ~0.397$ & $       ~0.005$ & $
      -0.476$ & $       ~0.010$ & $       ~0.079$ & $       0.007$ \\
$145.6$ & $       0.223$ & $       ~0.006$ & $       ~0.322$ & $       ~0.006$ & $
      -0.558$ & $       ~0.008$ & $       ~0.198$ & $       0.007$ \\
$148.0$ & $       0.227$ & $       ~0.005$ & $       ~0.279$ & $       ~0.005$ & $
      -0.552$ & $       ~0.007$ & $       ~0.255$ & $       0.006$ \\
$150.5$ & $       0.234$ & $       ~0.005$ & $       ~0.248$ & $       ~0.005$ & $
      -0.528$ & $       ~0.007$ & $       ~0.302$ & $       0.006$ \\
$153.4$ & $       0.219$ & $       ~0.006$ & $       ~0.198$ & $       ~0.007$ & $
      -0.456$ & $       ~0.009$ & $       ~0.343$ & $       0.006$ \\
$155.9$ & $       0.203$ & $       ~0.006$ & $       ~0.176$ & $       ~0.006$ & $
      -0.370$ & $       ~0.008$ & $       ~0.367$ & $       0.005$ \\
$158.3$ & $       0.171$ & $       ~0.005$ & $       ~0.137$ & $       ~0.006$ & $
      -0.310$ & $       ~0.007$ & $       ~0.372$ & $       0.005$ \\
$160.2$ & $       0.166$ & $       ~0.007$ & $       ~0.129$ & $       ~0.009$ & $
      -0.279$ & $       ~0.008$ & $       ~0.365$ & $       0.008$ \\
$163.1$ & $       0.127$ & $       ~0.006$ & $       ~0.122$ & $       ~0.008$ & $
      -0.192$ & $       ~0.006$ & $       ~0.329$ & $       0.007$ \\
$166.0$ & $       0.106$ & $       ~0.005$ & $       ~0.110$ & $       ~0.007$ & $
      -0.100$ & $       ~0.005$ & $       ~0.284$ & $       0.006$ \\
$168.8$ & $       0.078$ & $       ~0.005$ & $       ~0.101$ & $       ~0.007$ & $
      -0.032$ & $       ~0.005$ & $       ~0.226$ & $       0.006$ \\
$172.8$ & $       0.023$ & $       ~0.012$ & $       ~0.103$ & $       ~0.013$ & $
       0.046$ & $       ~0.005$ & $      $ & $            $ \\
$174.8$ & $       0.020$ & $       ~0.009$ & $       ~0.090$ & $       ~0.011$ & $
       0.052$ & $       ~0.004$ & $      $ & $            $ \\
$176.9$ & $       0.015$ & $       ~0.008$ & $       ~0.088$ & $       ~0.010$ & $
       0.077$ & $       ~0.004$ & $      $ & $            $ \\
$179.0$ & $       0.002$ & $       ~0.011$ & $       ~0.085$ & $       ~0.012$ & $
       0.087$ & $       ~0.005$ & $      $ & $            $ \\
\hline \hline
\end{tabular}
\caption{Data table for 
 analyzing powers for $d$--$p$ elastic scattering at $70~\rm MeV/u$ 
 measured with the SMART system.
 \label{tab140_s}}
\end{table}
\end{document}